\documentclass[11pt]{article}
\usepackage[a4paper,margin=1in]{geometry}
\usepackage{graphicx}
\usepackage{amsmath,amssymb}
\usepackage[numbers,sort&compress]{natbib}
\usepackage{hyperref}
\usepackage{authblk}
\usepackage[percent]{overpic}

\usepackage{siunitx}
\usepackage{siunitx}
\date{}

\begin{document}

\title{Aerosol memory in stratocumulus clouds leads to noise-induced patterns and non-ergodic sampling}

\author[1]{Benjamin Hernandez}
\author[1,2]{Franziska Glassmeier}

\affil[1]{Department of Geosciences and Remote Sensing, Delft University of Technology, Delft, Netherlands}
\affil[2]{Max Planck Institute for Meteorology, Hamburg, Germany}

\maketitle

\begin{abstract}

Stratocumulus cloud decks exhibit bistability between patterns of high (closed cells) and low (open cells) cloud fraction. Localized transitions between these two states (pockets of open cells) have been observed  but their underlying mechanism remains unclear.
We model stratocumulus and their interaction with atmospheric aerosol as a data-driven and physics-informed stochastic dynamical system with time-dependent parameters.
This allows us to show that pockets of open cells result from noise-induced transitions between the stratocumulus patterns.
We find comparable timescales for these transitions, mesoscale self-organization into patterns and the evolution of large-scale parameters.
This lack of timescale separation corresponds to an aerosol memory in cloud evolution and means that the
sampling of stratocumulus states by polar-orbiting satellites lacks the encoding of process information that would be present for an asymptotic and ergodic sampling.

\end{abstract}

\section{Introduction}

Clouds exert a notable net cooling effect on the climate system and it remains a key uncertainty of climate projections how this cooling effect changes \cite{bellouinBoundingGlobalAerosol2020,sherwoodAssessmentEarthsClimate2020}.
The processes that give rise to this uncertainty cover a broad range of scales --- from the formation of cloud droplets on atmospheric aerosol particles to their interaction with planetary-scale circulations. Linking these scales, the atmospheric \emph{mesoscale} is characterized by the organization of cloud fields across hundreds of kilometers, evolving over many hours.
Mesoscale organization is particularly relevant and striking in stratocumulus cloud decks.
Stratocumulus cover 20\% of the global oceans and contribute 50\% to the overall net cooling effect of clouds \cite{woodStratocumulusClouds2012,lecuyerReassessingEffectCloud2019}.
Subtropical stratocumulus feature bi-stability between two distinct types of cellular patterns (Fig:~\ref{fig:1}): a \emph{closed-cell} pattern, where cloudy cell centers are surrounded by a cloud-free outline, and an inverse pattern of \emph{open cells}. The closed-cell state has higher cloud fraction than the open-cell state.

As first suggested Baker and Charlson in 1990~\cite{bakerBistabilityCCNConcentrations1990}, hereafter BC90, and later confirmed by observations and detailed modeling studies, the self-organization of stratocumulus into the two patterns is driven by feedback processes in rain formation \cite{stevensEvaluationLargeEddySimulations2005,
Petters_2006,
sharonAerosolCloudMicrophysical2006,
savic-jovcicStructureMesoscaleOrganization2008, xueAerosolEffectsClouds2008,
KazilWangFeingold2011,
WoodBrethertonLeon2011,
BernerBrethertonWood13,
feingoldReversibilityTransitionsClosed2015}.
The colloidal instability of rain formation requires sufficiently large cloud droplets.
Cloud droplet size grows with the depth of a cloud and is modulated by the aerosol conditions, which determine the number concentration of droplets.
The open-cell pattern is thus favored by low aerosol abundance \cite{yamaguchiStratocumulusCumulusTransition2017} and large-scale conditions that favor deeper clouds \cite{hoffmannLiquidWaterPath2020}, i.e., deeper boundary layers \cite{woodSpatialVariabilityLiquid2006}, increased stability and increased surface fluxes \cite{McCoy_2017, gryspeerdtObservingShorttimescaleCloud2022}.
Stratocumulus decks more frequently encounter such conditions as they are advected from their formation regions close to the west coast of the continents over increasingly high sea-surface temperatures \cite{Muhlbauer2014,watson-parrisLargeScaleAnalysisPockets2021}.

It proves difficult, however, to link the formation of \emph{pockets of open cells} (Fig.~\ref{fig:1}) that are embedded within an otherwise closed-cell stratocumulus deck \cite{BrethertonUttalFairall04,stevensEvaluationLargeEddySimulations2005} to
local variability in these large-scale controls \cite{ smalleyLagrangianAnalysisPockets2022}.
This lack of evidence that pockets of open cells are externally imprinted raises the question if internal variability within close-cell stratocumulus decks can explain open-cell formation.
In other words, pockets of open cells might be the result of \emph{noise-induced transitions} between the two stratocumulus patterns.
To test this hypothesis, we need to determine if cloud-scale fluctuations in boundary-layer dynamics are large enough to trigger transitions from the closed-cell to the open-cell state of stratocumulus on observable timescales.

We address this question by describing stratocumulus decks and their interaction with aerosols as a stochastic dynamical system with time-dependent parameters. Combining our process understanding with large-eddy-simulation data, we show that the equilibration of aerosol processes in closed cells is slow compared to the evolution of the large-scale conditions. This lack of timescale separation means that the evolution of the cloud system depends on the initial aerosol conditions. Taking this aerosol memory into account, we find that pockets of open cells can indeed be understood as noise-induced transitions.
Combined, aerosol memory and the rarity of noise-induced transitions cause the sampling of the aerosol-stratocumulus state space to deviate from its asymptotic distribution. This transient behavior, or \emph{non-ergodicity} \cite{feingoldOpinionInferringProcess2025}, needs to be considered to obtain process understanding from satellite snapshots.

\section{Aerosol-stratocumulus interactions as data-driven dynamical system with three effective scales}

\begin{figure}[t!]
\centering
\includegraphics[width=\linewidth]{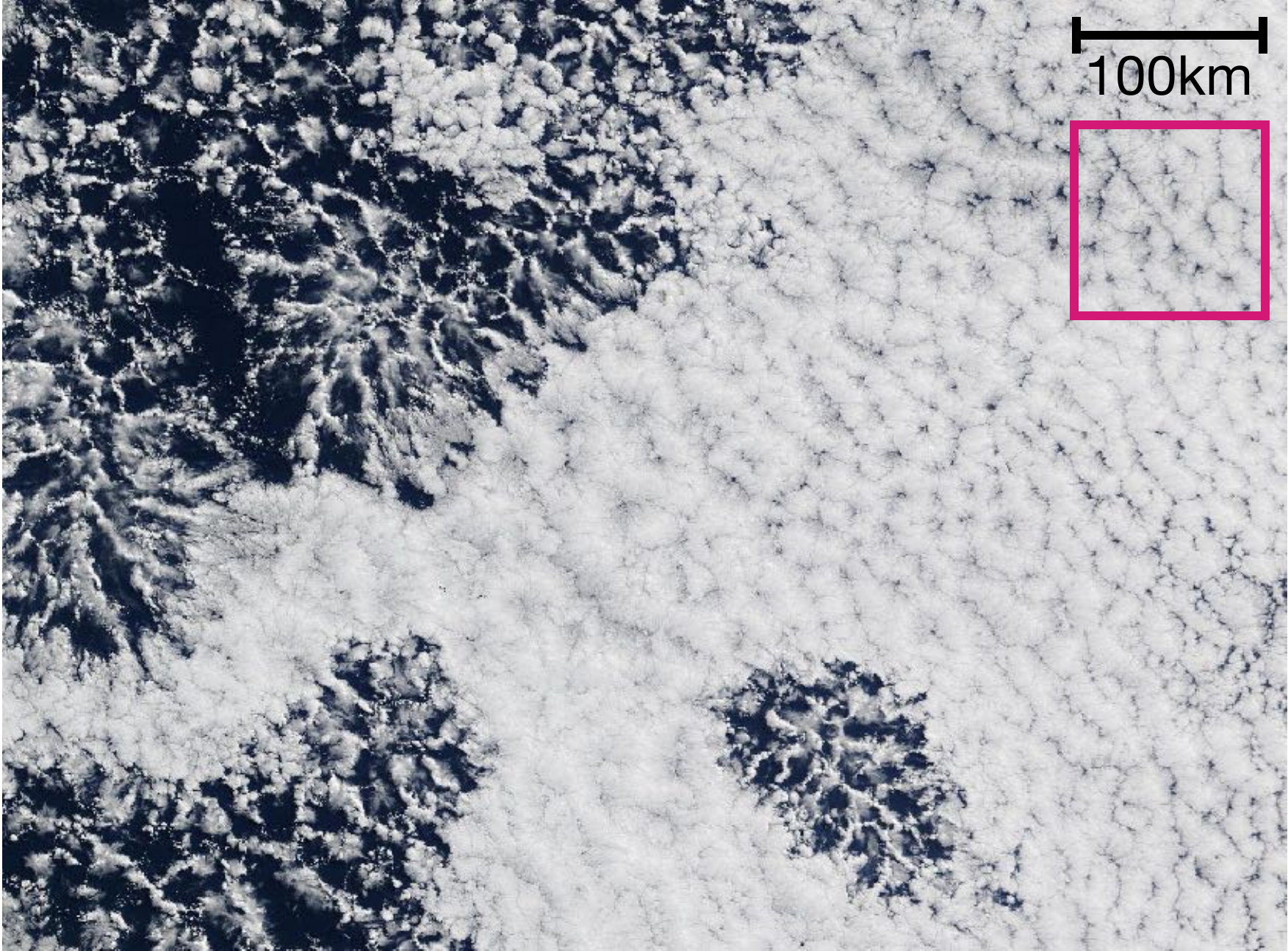}
\caption{\textbf{Stratocumulus cloud decks} feature bistability between closed cells (high cloud fraction) and open cells (low cloud fraction). The magenta box indicates the typical mesoscale size of a stratocumulus system as considered here. A pocket of open cells is shown at the lower right. From Jun 27, 2025, off the coast of Chile ($26^\circ \textrm{S}, 79^\circ \textrm{W}$).
We acknowledge the use of imagery from the Worldview Snapshots application, part of the Earth Science Data and Information System.}
\label{fig:1}
\end{figure}

Our description of stratocumulus focuses on their mesoscale self-organization into the closed- and open-cell states. Considering a mesoscale region, or cloud deck, as our system (Fig.~\ref{fig:1}), this evolution is effectively captured in a 2-dimensional state space of vertically integrated and spatially averaged condensate (liquid water path) $L$ and a mean cloud-droplet number concentration $N$ in the stratocumulus deck \cite{glassmeierAerosolcloudclimateCoolingOverestimated2021}.
Stratocumulus cloud fraction can be emulated as function of these two dimensions \cite{glassmeierEmulatorApproachStratocumulus2019}, $C = f_\textrm{C}(N,L)$, such that transitions in $N$-$L$ space imply distinct changes in cloud fraction as they distinguish open- and closed-cell stratocumulus.

We conceptualize the multiscale nature of the aerosol-stratocumulus system by three effective scales, where we couple the mesoscale dynamics (deterministic drift) to the smaller scales of cloud formation (stochastic diffusion) and to the large-scale drivers (time-dependent external parameters):
\begin{equation}
\label{eq:evolution}
\begin{split}
\frac{d}{dt}
\begin{pmatrix}
N\\
L
\end{pmatrix}
&=
\underbrace{
\begin{pmatrix}
-\beta N^2 + \dfrac{1}{z(t)}\!\left(S_0 - f_\mathrm{P}(N,L)\right)\\[1ex]
-\dfrac{L - \overline{L}(t)f_\textrm{L}(N)}{\tau(N)}
\end{pmatrix}}_{\text{drift}}\\[2ex]
&\quad +
\underbrace{
\epsilon
\begin{pmatrix}
1 & 0\\
\alpha & \sqrt{1-\alpha^2}
\end{pmatrix}
\begin{pmatrix}
\xi_N\\
\xi_L
\end{pmatrix}}_{\text{diffusion}}.
\end{split}
\end{equation}
Following BC90, the $N$-dynamic here balances the aerosol source --- mixed across the boundary-layer depth $z$ --- with sinks from coagulation ($\beta N^2$ term) and rain (rain formation function $f_\textrm{P}$). The mesoscale $L$-dynamic in Eq.~\ref{eq:evolution} relaxes with a timescale $\tau$ \cite{glassmeierAerosolcloudclimateCoolingOverestimated2021}, where a function $f_\textrm{L}(N)$ accounts for adjustments in the steady-state value of $L$ to the aerosol conditions \cite{glassmeierAerosolcloudclimateCoolingOverestimated2021, chenMagnitudeTimescaleLiquid2025, hoffmannProcessesDeterminingSlope2025}. Cloud-scale fluctuations caused by boundary-layer turbulence are modeled as white noise $\xi$ with amplitude $\epsilon$, which allows for correlation $\alpha$ (Appendix A3) between fluctuations in $N$ and $L$~\cite{hoffmannImpactAerosolCloud2024}.

The large-scale controls are parameterized by an aerosol-source parameter $S_0$, a large-scale meteorological control parameter $\overline{L}$, which approximately corresponds to the steady-state value of $L$ averaged over its $N$-dependence, and boundary-layer height $z$. We keep the aerosol source $S_0$ fixed, while the meteorological driver $\overline{L}(t)$ and the boundary-layer height $z(t)$ increase over time to capture the typical stratocumulus advection over higher sea-surface temperatures (Appendix, Fig.~A1). As $\overline{L}$ and $z$ co-vary, our analysis focuses on $\overline{L}$ as main time-dependent parameter. It is further modulated periodically by the diurnal cycle such that $\overline{L}$ is larger during the night than during the day, when short-wave radiative heating thins the cloud.

We obtain the quantitative values for the mesoscale dynamics and noise amplitude $\epsilon$ from an ensemble of large-eddy simulations from Glassmeier et al.~\cite{glassmeierEmulatorApproachStratocumulus2019}, GF19 hereafter, that explores the mesoscale state-space by varying initial conditions for $S_0$ and $\overline{L}$ fixed to typical conditions in the bistable region.
For the typical evolution $\overline{L}(t)$ and $z(t)$, we rely on a large-eddy simulation case study
by Yamaguchi et al.~\cite{yamaguchiStratocumulusCumulusTransition2017}.
In addition to providing an analytical form, the physics-informed approach Eq.~\ref{eq:evolution} allows us to extrapolate to values of the large-scale parameters for which no large-eddy simulations are available.

\subsection{Open- and closed-cells as co-existing steady states with hysteresis}
\begin{figure*}[t!]
\centering
\includegraphics[width=1\linewidth]{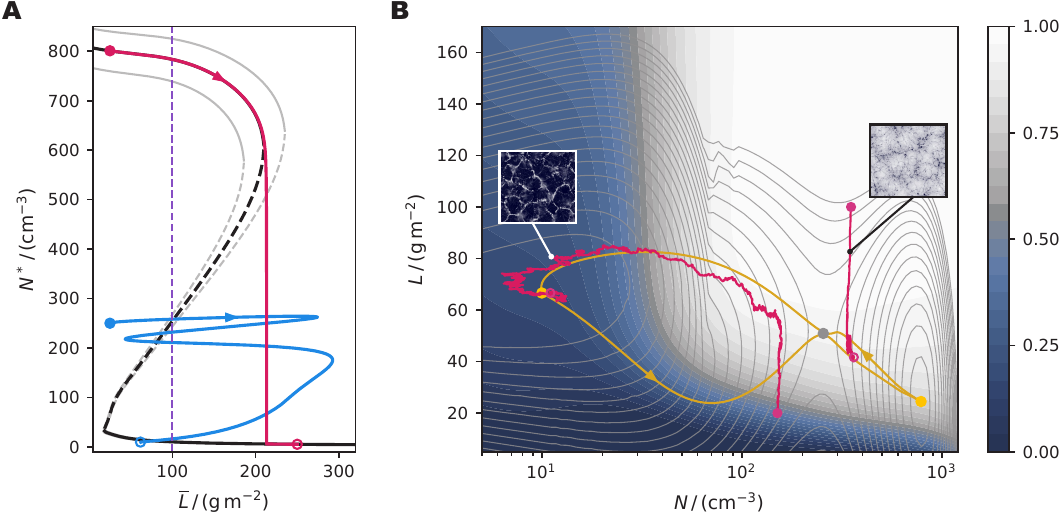}
\caption{\textbf{Three-scale effective model of aerosol-stratocumulus interactions Eq.~\ref{eq:evolution}.} (A) Steady-state value $N^*$ of cloud-droplet number $N$ as function of the meteorological control parameter $\overline{L}$ for aerosol background values $S_0$ as in GF19, with $S_0 \pm 10\%$ indicated in light gray.
Solid and dashed black lines represent stable and unstable steady states, respectively. The blue curve illustrates a two-day stratocumulus evolution forced by $\overline{L}(t)$ increasing with added diurnal oscillations.
The magenta curve shows a classical tipping trajectory under a quasi-steady increase with $\textrm{d}\overline{L}(t)\approx 0$. Filled circles mark initial conditions, open circles the final states. The dashed vertical line in purple indicates the GF19 reference value of $\overline{L}$. (B) Freidlin-Wentzell quasi-potential (grey contour lines) representing the $N$-$L$ evolution for reference values of the large-scale parameters (purple line in panel A) and cloud fraction (color contours). The potential minima correspond to the two stable states indicated by the intersection of the purple dashed line in panel A with the $N^*$-curve. Insets illustrate the open- (low $N$) and closed-cell (high $N$) pattern of large-eddy simulations corresponding to the steady states. Jagged magenta lines show exemplary trajectories of individual solutions to Eq.~\ref{eq:evolution}. Yellow lines show instantons, i.e., the most likely paths for noise-induced transitions.}
\label{fig:2}
\end{figure*}

Figure~\ref{fig:2} illustrates the behavior of Eq.~\ref{eq:evolution}.
The key characteristics of the mesoscale evolution are captured by its quasi-potential (Fig.~\ref{fig:2}B), which generalizes the concept of a potential landscape to non-gradient systems (Appendix A1, A2). It represents the system's stability structure and response to noise~\cite{freidlinRandomPerturbationsDynamical2012}.
Its two minima correspond to the co-existence of two steady-states, i.e., the bistability of the system.
Comparison to the spatial structure of corresponding large-eddy simulations lets us identify these two steady states with the open- and closed-cell state of stratocumulus and their distinct differences in cloud fraction (Fig.~\ref{fig:2}B, insets).
The closed- and open-cell steady states balance the aerosol source in Eq.~\ref{eq:evolution} with the coagulation and precipitation sink, respectively.
Which of the two mesoscale steady states a system evolves two depends on which basin of the potential landscape the initial state lies in (Fig.~\ref{fig:2}B, magenta).
We note that the basin boundary follows the onset of rain formation~\cite{glassmeierAerosolcloudclimateCoolingOverestimated2021} and is not aligned with cloud fraction such that equilibration to the open-cell steady state may start from high cloud fraction.
We will refer to such evolution as a \emph{equilibration} in contrast to \emph{transitions} between the open- and closed-cell basins.

As illustrated by~BC90, a saturation of the collision–coalescence efficiency at low aerosol concentrations in the rain formation function $f_\textrm{P}$ in Eq.~\ref{eq:evolution} leads to saddle-node bifurcations to and from bi-stability in both large-scale parameters (Fig.~\ref{fig:2}A; Appendix, Fig.~A2).
Starting from a closed-cell steady state, the transition to open cells occurs once $\overline{L}(t)=L_c$ passes the threshold value where the closed-cell branch ceases to exist (Fig.~\ref{fig:2}A, magenta), following classic tipping-point behavior~\cite{ashwinTippingPointsOpen2012}.
A decrease of $\overline{L} < L_c$ that is small compared to the extend of the bi-stable region would not lead to recovery of the closed cells state, i.e., the rain-formation feedback locks the system in the low-$N$ state and causes the hysteretic behavior characteristic of tipping points.

\subsection{Aerosol dynamics cause lack of timescale separation}
The gradient of the quasi-potential captures the timescale of equilibration into the states for small non-gradient contributions to the evolution as in our case (Fig.~\ref{fig:2}B, gray contours, note the logarithmic scale). While the open-cell state is approached with a timescale of $4\,$h, the closed-cell basin is shallower and wider with an equilibration timescale of $10\,$h in the L- and multiple days in the N direction.
These different timescales match different physical processes, namely the rain-formation feedback for the open cells~\cite{sharonAerosolCloudMicrophysical2006,savic-jovcicStructureMesoscaleOrganization2008, wangModelingMesoscaleCellular2009, yamaguchiRelationshipOpenCellular2015}, the balancing of cloud-top radiative cooling and entrainment for the closed-cell $L$-dynamics~\cite{lillyModelsCloudtoppedMixed1968, schubertMarineStratocumulusConvection1979, hoffmannLiquidWaterPath2020}, and the slow accumulation of aerosols in the boundary layer due to the source $S_0$ in the $N$-dynamics in Eq.~\ref{eq:evolution} \cite{KazilWangFeingold2011}.
This illustrates that our conceptual mesoscale corresponds to a range of timescales.
Comparison to the timescale of the large-scale parameter $\overline{L}(t)$, which amounts to days, and the timescale of cloud-formation of minutes to an hour, furthermore shows that the mesoscale seamlessly couples to and in between these scales.
The aerosol dynamics considered in Eq.~\ref{eq:evolution} thus prevents a separation of the timescales of thermodynamics and boundary-layer evolution as discussed in~\cite{brethertonSlowManifoldsMultiple2010}.

\subsection{Lack of timescale separation amplifies hysteretic memory and facilitates transition} This lack of timescale separation, together with the large size of the closed-cell basin, implies that closed-cell stratocumulus are unlikely to be equilibrated to their steady state. Consequently, their observed state in $N$-$L$-space is not well approximated by steady-state values, as often assumed in conceptual models~\cite{hoffmannImpactAerosolCloud2024, korenAerosolCloudPrecipitation2011}, and the evolution cannot be reduced to an effectively 1-dimensional description.
This transient nature affects how the system tips into the open-cell state.
Tipping is typically described for systems that remain in steady state, i.e., large-scale parameters change on timescales that are clearly separated from the dynamics of the system (Fig.~\ref{fig:2}A, magenta trajectory).
When the large-scale conditions $\overline{L}(t)$ change on a shorter timescale, notably due to the diurnal cycle, the non-equilibrated system can cross the instable equilibrium line earlier, i.e., for $\overline{L}(t) < L_\textrm{c}$ (Fig.~\ref{fig:2}A, blue trajectory).
This behavior bears similarity to rate-induced tipping~\cite{ritchieRateinducedTippingNatural2023} in that transitions occur without the loss of a stable equilibrium, but it differs in that it is not driven by a critical forcing rate or basin instability but by what might be considered an amplified hysteretic \emph{memory} effect: the system's evolution not only depends on the initial basin of attraction as for hysteretic systems with timescale separation but on the specific initial value of $N$. Note that the system's evolution is nevertheless fully characterized by its current state in $N$-$L$ space, i.e., the 2-dimensional system does not feature memory in the sense of a non-Markovian evolution, which would require knowledge of its state at previous timesteps.

The diurnal cycle also drives decreases in $\overline{L}$.
As its amplitude is comparable to the width of the bi-stable region, it transitions the system back to the closed-basin of attraction for intermediate $N$ and weak rain but not for low $N$ and strong rain-formation feedback (Fig.~\ref{fig:2}A, blue trajectory).

\section{Pockets of open cells as memory-enabled, noise-induced transitions}

In addition to changes in the large-scale drivers of $\overline{L}(t)$, transitions can also be triggered by fluctuations.
The probability of such noise-induced transitions is controlled by the stability of the basin of attraction, quantified by the Freidlin--Wentzell quasi-potential $U$ (Fig.~\ref{fig:1}B). It translates into a mean transition time,
$\mathcal{T} \propto \exp(\Delta U/\epsilon^2)$, where $\Delta U$ denotes the quasi-potential barrier between the initial and final state and $\epsilon$ the noise amplitude in Eq~\ref{eq:evolution}. The quasi-potential landscape and thus transition statistics depend on the large-scale parameter values (Appendix, Fig.~A3).
For equilibrated systems deep within a basin, i.e., under effectively fixed large-scale conditions, $\overline{L}\approx\textrm{const}$, we obtain mean exit times larger than $100\,$days (Appendix, Fig.~A3). This notably exceeds the day-long lifetime of stratocumulus decks.
Even given how rarely they are observed \cite{watson-parrisLargeScaleAnalysisPockets2021, smalleyLagrangianAnalysisPockets2022}, such noise-induced transitions are thus too unlikely to explain the formation of pockets of open cells.

\begin{figure}[t!]
\centering
\begin{minipage}{0.48\textwidth}
    \begin{overpic}[width=\linewidth]{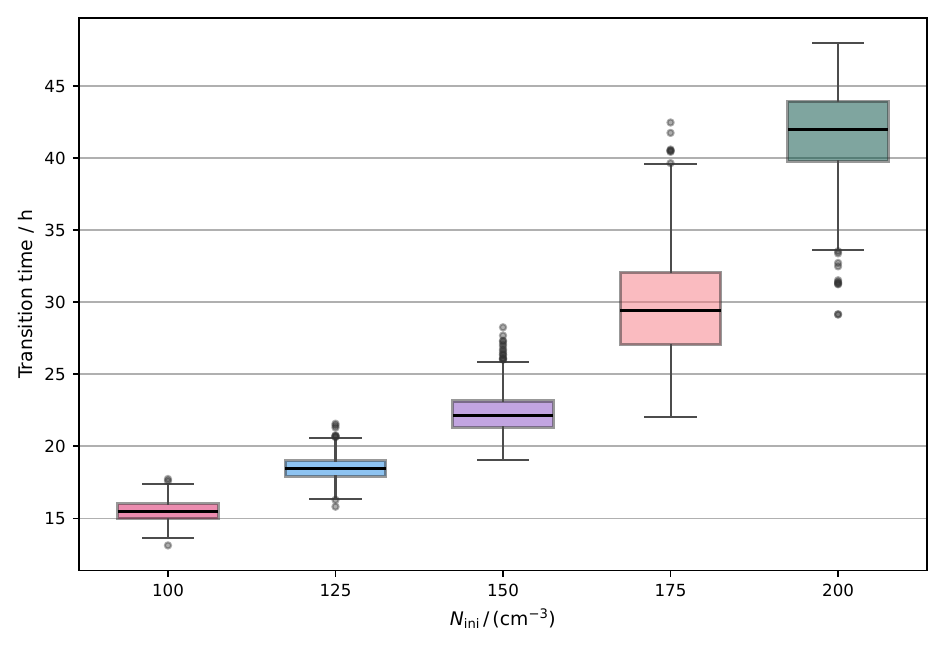}
        \put(3,70){\large\textsf{\textbf{A}}}
    \end{overpic}
\end{minipage}
\hfill
\begin{minipage}{0.48\textwidth}
    \begin{overpic}[width=\linewidth]{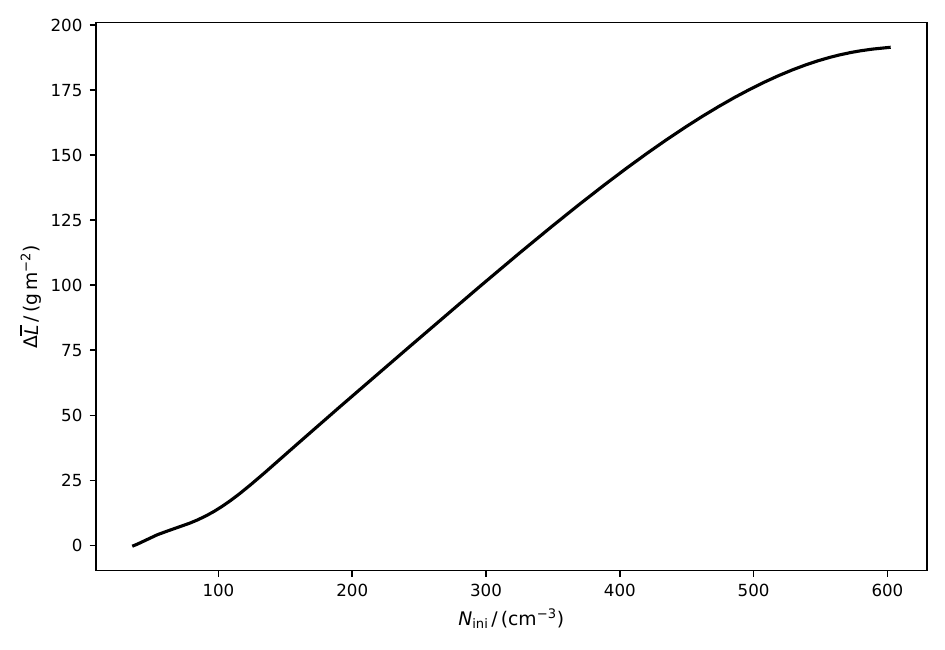}
        \put(2,70){\large\textsf{\textbf{B}}}
    \end{overpic}
\end{minipage}
\caption{\textbf{Noise-induced transitions from closed to open cells under time-dependent forcing.} (A) Transition time as function of initial cloud droplet number $N_\mathrm{ini}$. For each value of $N_{\mathrm{ini}}$, we evolve an ensemble of $n=500$ stochastic trajectories according to Eq.~\ref{eq:evolution}, initialized with $N_\textrm{ini}$ and
$L_{\mathrm{ini}}=\overline{L}(t=0)f_\textrm{L}(N_\textrm{ini})$. All other parameters are set to the GF19 configuration. Transition times are defined as the first time at which cloud fraction drops below 0.4.
(B) Effective width $\Delta\overline{L}$ of the bistable region of $\overline{L}$ as a function of the initial condition $N_\mathrm{ini}$. The effective width $\Delta\overline{L}$ is defined as difference between the value of $\overline{L}$ at the bifurcation point on the open-cell branch ($\overline{L}\approx20\,\textrm{g}\,\textrm{m}^{-2}$ in Fig.~\ref{fig:2}) and the value of $\overline{L}$ where the unstable branch intersects $N^*=N_\textrm{ini}$ ($\overline{L}\approx100\,\textrm{g}\,\textrm{m}^{-2}$ in Fig.~\ref{fig:2}). For larger $N_{\mathrm{ini}}$ than shown, the effective width saturates at the maximum value displayed.
}
\label{fig:3}
\end{figure}

In contrast, systems subject to an evolving large-scale parameter $\overline{L}(t)$ spend considerable time close to the basin boundary (Fig.~\ref{fig:2}A, blue trajectory).
This enables the same noise levels that are too low to affect equilibrated systems to advance or delay transition times relative to the mean value (Fig.~\ref{fig:3}A).
We identify pockets of open cells as the outliers in the distribution of transition times that, due to noise, transition substantially earlier than the mean stratocumulus deck. Larger values of $N_{\textrm{ini}}$ correspond to an effectively larger bi-stable region, i.e., a larger range of $\bar{L}$ for which systems are susceptible to noise-induced transitions (Fig.~\ref{fig:3}B).
Larger values of $N_{\textrm{ini}}$ increase the spread of transition times (Fig.~\ref{fig:3}A), and thus the probability to form pockets of open cells. We note that larger $N$ also corresponds to brighter cloud decks, in which a stronger contrast in brightness makes pockets of open cells easier to detect.

The spread of transition rates is modulated by the correlation structure of the noise, where anti-correlated noise leads to larger transition-time variability.
Perturbations that increase $N$ and decrease $L$, or vice versa, are thus more likely to form pockets of open cells (Appendix, Fig.~A5).
The sensitivity is even more pronounced in a static landscape without large-scale evolution, where anti-correlated fluctuations result in mean exit times that are orders of magnitude smaller than correlated fluctuations (Appendix, Fig.~A4). We argue that this sensitivity arises from the geometry of the most probable exit pathways (instantons), which follow trajectories with negative slope in $N$--$L$ space (Fig.~\ref{fig:2}B, yellow). Anti-correlated fluctuations therefore project efficiently onto this preferred escape direction and enhance early transitions, whereas positively correlated fluctuations hinder them. We note that instantons exhibit a clear directionality such that optimal pathways into and out of the steady states do not coincide. This directionality reflects the non-equilibrium nature of stratocumulus and corresponding entropy production, which is mostly driven by moist processes~\cite{HernandezSinghYamaguchi26}.

\section{Snapshot sampling of stratocumulus is non-ergodic}

Consistent with the generally low observed frequency of pockets of open cells~\cite{watson-parrisLargeScaleAnalysisPockets2021, smalleyLagrangianAnalysisPockets2022}, noise-induced transitions remain rare in all scenarios.
This has implications for the sampling of stratocumulus states by snapshots from polar-orbiting satellites (Fig.~\ref{fig:4}).
Noise-induced transitions ensure that both basins of attraction are sampled if there is a timescale separation between the deterministic dynamics of the system and the evolution of the large-scale parameters such that there is enough time for even improbable transitions to occur frequently.
The resulting asymptotic evolution of the system explores its state-space with a distribution that is given by the shape of the quasi-potential (Appendix A1) because systems spend more time in the steady-state regions, where they persist, than in transient regions, which they rapidly evolve from~\cite{freidlinRandomPerturbationsDynamical2012}.
The residence time of the system in these different regions corresponds to the probability to observe the system in this state. The sampling frequency of the system state by independent snapshots of its asymptotic evolution is thus the same as the asymptotic distribution of an individual system.
This situation can be described as ergodic in the sense that the distribution of an ensemble of snapshots is equivalent to the distribution that would be obtained from an infinitely long timeseries of a single system \cite{feingoldOpinionInferringProcess2025}.

A lack of timescale separation between the large-scale and mesoscale dynamics means that the timescale of noise-induced transitions is so long that the system cannot explore both basins of attraction --- the system is stuck in its initial basin. For stratocumulus, this is reflected in closed cells being more frequent than open cells \cite{Muhlbauer2014}. In addition, a lack of timescale separation means that the locations of the steady states change while the system is still equilibrating --- the system chases a moving target. As a consequence, the closed-cell steady state is practically unobservable.

Both effects mean that the observed sampling is not directly related to the shape of a quasi-potential.
As the shape of the quasi-potential reflects the physical processes that shape it, such a snapshot sampling does not directly correspond to information about processes.
Instead, the sampling also depends on the initial distribution of the ensemble in $N$-$L$ space (Fig.~\ref{fig:4}).
The relative importance of initial distribution vs process information as captured by the quasi-potential is determined by the timescales of noise-induced transitions, mesoscale self-organization and large-scale evolution.
Inferring process information from snapshot data as discussed in \cite{feingoldOpinionInferringProcess2025} therefore requires accounting for these timescales and the system's memory of the initial distribution.

\section{Discussion and conclusions}

\begin{figure*}[t!]
\centering
\includegraphics[width=\textwidth]{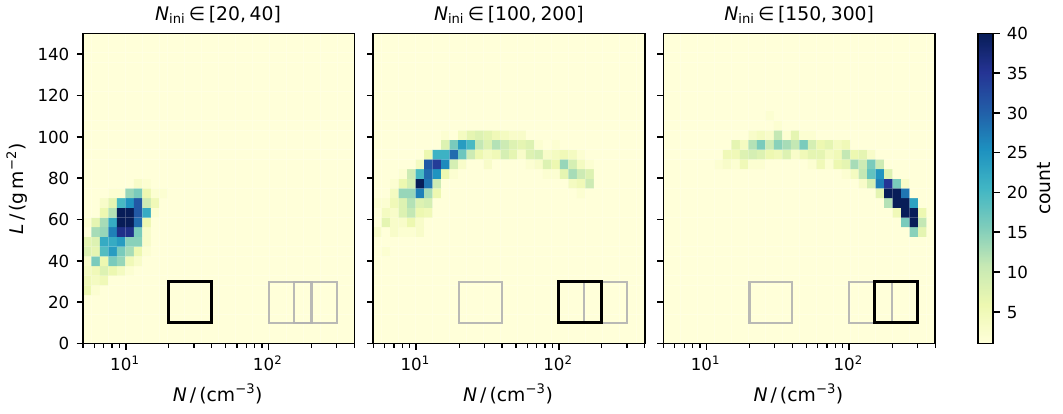}
\caption{\textbf{Snapshot sampling for different initial conditions.} Three ensembles of $n=1000$ stochastic trajectories are evolved according to Eq.~\ref{eq:evolution} under identical, time-varying large-scale forcing $\overline{L}(t)$ but with different initial-condition ranges, progressively shifted toward higher $N_{\mathrm{ini}}$ (left to right). Dark boxes indicate the initial-condition range used for each ensemble, while light boxes show the ranges used in the other experiments. Colors indicate the sampling of the $N$-$L$ state space by snapshots of the ensemble at time $t=24\,\mathrm{h}$.
}
\label{fig:4}
\end{figure*}

Our results highlight the role of aerosol dynamics for stratocumulus evolution.
While aerosols and the microphysical cloud state they set are, in general, second order relative to dynamic and thermodynamic cloud controls, aerosols matter because they evolve on timescales that are comparable or slower than the dominant cloud processes and thus give memory to the system.
The aerosol timescale in the model considered here (Eq.~\ref{eq:evolution}, Fig.~\ref{fig:2}) is largely set by the aerosol source $S_0$. It parametrizes aerosol processes in the GF19 large-eddy simulations. These are simplified in that they do not allow for interactive emissions of aerosol at the surface, and that they do not consider the free troposphere as an aerosol reservoir. Due to wind-driven aerosol emissions in the open-cell state \cite{KazilWangFeingold2011, gryspeerdtObservingShorttimescaleCloud2022}, the former is expected to slow down the rain feedback. The latter will speed up equilibration at very high $N$ due to detrainment as aerosol sink \cite{KazilWangFeingold2011, BernerBrethertonWood13}. This additional complexity would thus even further reduce timescale differences among processes.
As computational resources allow for larger domains and longer simulations, capturing such complex aerosol processes and their mesoscale memory becomes increasingly important.

As one consequence of mesoscale aerosol dynamics and the associated lack of timescale separation, we have shown that pockets of open cells are a form of noise-induced transitions, enabled by a lack of equilibration of the aerosol dynamics (Fig.~\ref{fig:3}).
In the model~Eq.~\ref{eq:evolution}, pockets of open cells are captured in a mesoscale-mean sense and correspond to temporal fluctuations only.
We do not account for their spatial structure (Fig.~\ref{fig:1}). As spatial interaction between neighboring open cells influence their evolution \cite{yamaguchiRelationshipOpenCellular2015}, we speculate that such interactions affect the cloud-scale fluctuations in Eq.~\ref{eq:evolution}. These interactions have been shown to be local \cite{glassmeierNetworkApproachPatterns2017}, however, justifying our spatially-agnostic approach.

The aerosol-induced lack of timescale separation also means that the initial aerosol condition affects observed stratocumulus states and transitions. This aerosol memory weakens the relationship between snapshot statistics and physical processes (Fig.~\ref{fig:4}). This notably affects the interpretation of the latest generation of snapshot observations that is currently becoming available and provides unprecedented detail \cite{Illingworth_2015}.
The multiscale perspective on aerosol-stratocumulus interactions discussed here, showcases the interplay of information about processes with timescales and memory.

In conclusion, the mesoscale and variables that evolve on it like aerosols need to be considered in the next generation of simulations and to deduce process understanding from the latest observations.
More broadly, aerosol-stratocumulus interactions, and the atmosphere in general, are a complex, multiscale system which cannot be understood by assuming timescale separations.
Despite their rarity and corresponding negligible effect on the global energy balance \cite{watson-parrisLargeScaleAnalysisPockets2021}, pockets of open cells are a fascinating window into this complexity.

\newpage
\section*{Aknowledgements}
We thank Robert Wood for insightful input  on~BC90 and Jürgen Kurths for helpful discussion of stochastic fluctuations. We are grateful to Takanobu Yamaguchi and Graham Feingold for the GF19 simulations. BH and FG acknowledge support from The Branco Weiss Fellowship ‐ Society in Science, administered by ETH Zurich. FG also acknowledges support by the European Union (ERC, MesoClou, 101117462). Views and opinions expressed are however those of the authors only and do not necessarily reflect those of the European Union or the European Research Council Executive Agency. Neither the European Union nor the granting authority can be held responsible for them.

\section*{Appendix}
\renewcommand{\thefigure}{A\arabic{figure}}
\setcounter{figure}{0}
\renewcommand{\thesubsection}{A\arabic{subsection}}
\setcounter{subsection}{0}

\subsection{Large-deviation theory}
For a stochastic system of the form of Eq.~\ref{eq:evolution}, one can define the `energy' barrier between the states $x$ and $x_0$ that needs to be overcome for an effective transition by minimizing the Freidlin and Wentzell action $S_T$~\cite{freidlinRandomPerturbationsDynamical2012},
\begin{equation}
\label{eq:Freidlin_Wentzell}
U(x) = \lim_{T\to\infty} \min_{\phi_{x,x_0}} S_T[\phi] = \lim_{T\to\infty} \min_{\phi_{x,x_0}} \frac{1}{2}\left\{ \int_0^T \left\| \dot{\mathbf{\phi}} - \mathbf{F}(\mathbf{\phi}) \right\|^2 dt \right\}
\end{equation}
where the minimum is to be taken between all the paths $\mathbf{\phi}$ that connect the two states in a time $T$ ($\phi(0) = x_0$, $\phi(T)=x$). This effectively defines the function $U$, the \textit{quasi-potential}, as a measure of deviation from the deterministic trajectory driven by the field $\mathbf{F}(\mathbf{x})$ (drift term in Eq.~\ref{eq:evolution}) when escaping from the attraction basin due to the random fluctuations.

The minimizing path of Eq.~\ref{eq:Freidlin_Wentzell} corresponds to the most probable escape trajectory in the weak-noise limit and is commonly referred to as the \emph{instanton}~\cite{zhouQuasipotentialLandscapeComplex2012}.
While individual stochastic realizations may deviate from this trajectory, large deviation theory implies that the probability of observing an escape path that remains within a distance $\epsilon$ of the instanton scales as
\begin{equation}
P\!\left(\left\|\mathbf{x} - \mathbf{x}_{\mathrm{inst}}\right\| < \epsilon\right)
\propto \exp\!\left(-\frac{U}{\epsilon^2}\right),
\end{equation}
demonstrating that escape events between the two states are exponentially concentrated around the instanton in the small-noise limit.

Finally, in the small noise limit, the quasi-potential is related to the asymptotic steady-state distribution $P(x)$ by the following scaling relationship~\cite{kikuchiRitzMethodTransition2020},
\begin{equation}
    P(x) \propto\exp\!\left({-\frac{U(x)}{\epsilon^2}}\right).
\end{equation}

\subsection{Quasi-potential and instanton computation}
In order to compute the quasi-potential, we used the PyRitz python library~\cite{kikuchiRitzMethodTransition2020}. The library employs a Ritz method (direct method) for the minimization of the Freidlin and Wentzell action functional (main text Eq.~10 of the main text). The paths over which the minimization takes place are approximated by Chebyshev polynomials, therefore the minimization becomes a simple non-linear optimization over the coefficients of the expansion, performed by the NLopt library using the SLSQP algorithm~\cite{johnsonNLoptNonlinearoptimizationPackage2007}. After dividing the state-space in an $n \times n$ grid, we compute the instanton $\psi$ and the respective value of the action $S_T$ for the paths starting on the first attractor $x_{a}$ and ending on the grid points $x_{i}$,
\begin{equation}
U_{a}(x_i) = \lim_{T\to\infty} \min_{\phi_{a,i}} S_T[\phi] = S[\psi_{a,i}].
\end{equation}

This gives us the quasi-potential for the first attractor. We then repeat the same procedure for the second attractor, and connect together the two local functions by requiring that
\begin{equation}
    U_{a}(x_s) + C_a = U_b(x_s) + C_b,
\end{equation}
where $C_a, C_b$ are simple additive constants, and $x_s$ is the saddle with the lowest action value on the separatrix between the two basins. The global quasi-potential is then determined by
\begin{equation}
    U(x) = \min_j  \left\{ U_{j}(x) + C_j \right\}.
\end{equation}
The interpolation and quadrature order for the polynomial approximation used are set to $n_p=8$ and $n_q = 80$, respectively. The tolerance is set to $10^{-12}$.

\subsection{Noise correlations}
In order to simulate the impact of correlations between the two noise components $\xi_N, \xi_L$, we define the covariance matrix as
\begin{equation}
    \boldsymbol{C} = \epsilon^2
\begin{bmatrix}
1 & \alpha \\
\alpha & 1
\end{bmatrix},
\end{equation}
where $\epsilon$ represents the noise level, and $\alpha$ the correlation coefficient. Given our white, uncorrelated noise vector $\boldsymbol{\xi} = (\xi_L, \xi_N)$, we need to find a matrix $\boldsymbol{B}$ such that
\begin{equation}
    \boldsymbol{\eta} = \boldsymbol{B}\boldsymbol{\xi}
\end{equation}
is a vector comprised of two correlated, gaussian distributed variables. To do this, we leverage the Cholesky decomposition, which allows us to factor the correlation matrix $\boldsymbol{C}$ as
\begin{equation}
    \boldsymbol{C} = \boldsymbol{B}\boldsymbol{B}^T,
\end{equation}
with $\boldsymbol{B}$ a lower triangular matrix.
Given the low dimensionality, it is easy to show that in our case
\begin{equation}
    \boldsymbol{B} = \epsilon
\begin{bmatrix}
1 & 0 \\
\alpha &  \sqrt{(1-\alpha^2)}
\end{bmatrix},
\end{equation}
which, in the stochastic formulation, becomes our diffusion matrix,
\begin{equation}
\label{eq:stochastic_system}
\dot{\mathbf{x}} = \mathbf{F}(\mathbf{x}) + \boldsymbol{B} \boldsymbol{\xi},
\end{equation}
where $\mathbf{F}(\mathbf{x})$ and $\boldsymbol{B}$ represent the drift and diffusion terms in Eq.~1 of the main text.

\begin{figure}[t!]
\centering
\includegraphics[width=\textwidth]{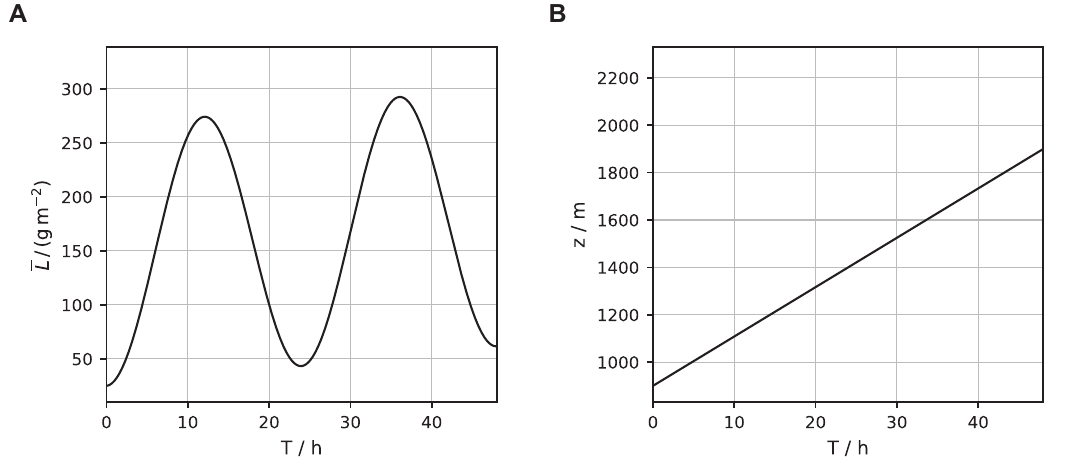}
\caption{Large-scale evolution of (A) the meteorological control parameter $\overline{L}(t)$ and (B) boundary-layer height $z(t)$ following the simulation in~\cite{yamaguchiStratocumulusCumulusTransition2017} that does not allow for drizzle formation.}
\end{figure}

\begin{figure}[t!]
\centering
\includegraphics[width=\textwidth]{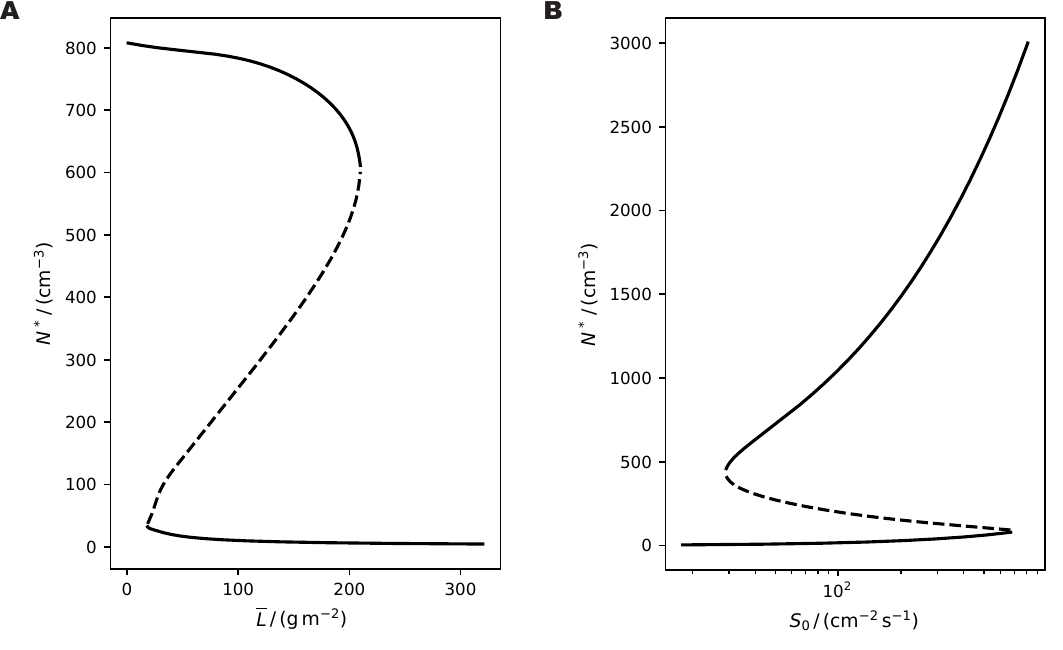}
\caption{Bifurcation plot of steady-state cloud-droplet numbr $N^*$ with respect to (A) the meteorological control parameter $\overline{L}$ and (B) the aerosol source parameter $S_0$. Solid lines represent the two stable states (open and closed cells for low and high $N^*$, respectively), while the dashed line shows the unstable steady-state branch. All other parameters are set to the GF19 configuration.}
\end{figure}

\begin{figure}[t!]
\centering
\includegraphics[width=\textwidth]{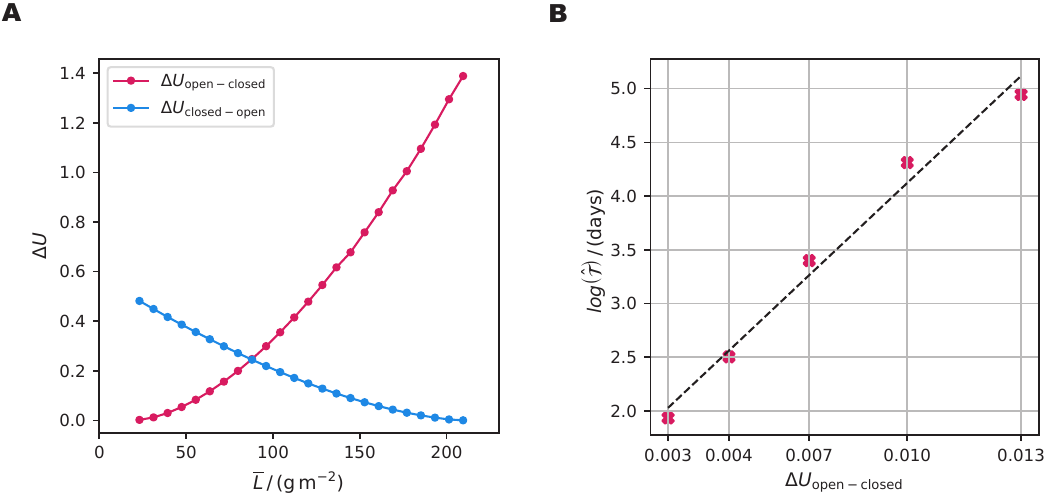}
\caption{Stability of the open- and closed-cell steady-state against perturbations. (A) Quasi-potential barriers $\Delta U$ between the open- and closed-cell states as a function of the bifurcation parameter $\overline{L}$ (similar behavior is obtained when varying $S_0$). (B) Scaling of mean exit time for noise-induced transitions with the quasi-potential barrier. Trajectories are initialized in the open-cell state for different values of $\overline{L}=[24,26,28,30,32]\,\mathrm{g\,m^{-2}}$. Each point is estimated from an ensemble of $n=200$ simulations. The noise level is increased to $6\%$ for computational efficiency. Quasi-potential values are multiplied by 100 for easier visualization. In both panels, all other parameters are fixed to the GF19 configuration.}
\end{figure}

\begin{figure}[t!]
\centering
\includegraphics[width=\textwidth]{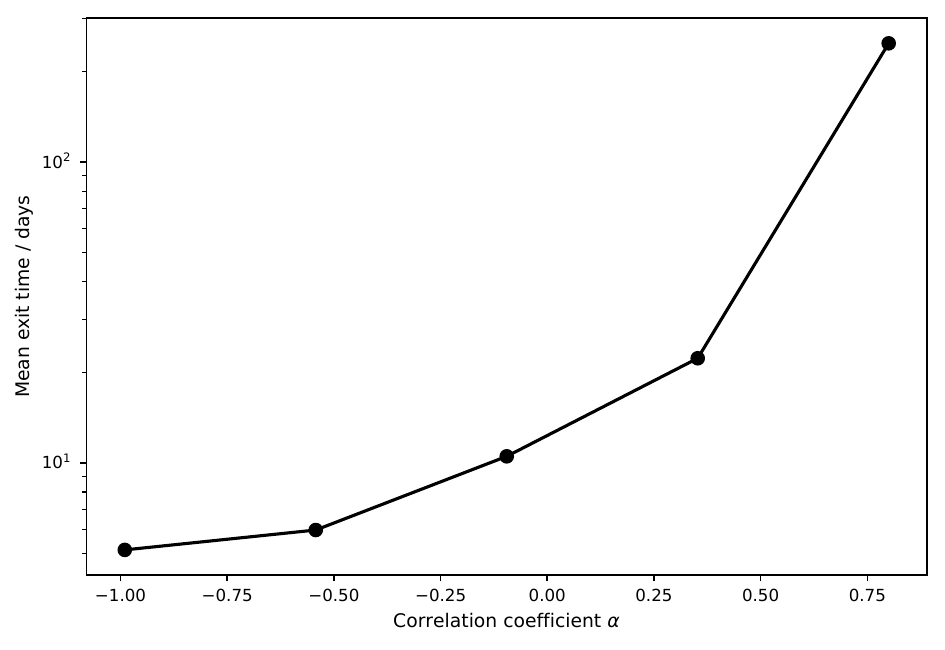}
\caption{Mean exit time for a static landscape, i.e. for fixed $\overline{L}=22\,\mathrm{g\,m^{-2}}$, as function of noise correlation $\alpha$. Each dot represents an ensemble of $n=200$ simulations initialized on the open-cell state. Exit time is defined as the first time at which a trajectory crosses the saddle. Simulations are run until all ensemble members have transitioned. The time step is fixed to $dt=60\,\mathrm{s}$. All other parameters are set to the GF19 configuration.}
\end{figure}

\begin{figure}[t!]
\centering
\includegraphics[width=\textwidth]{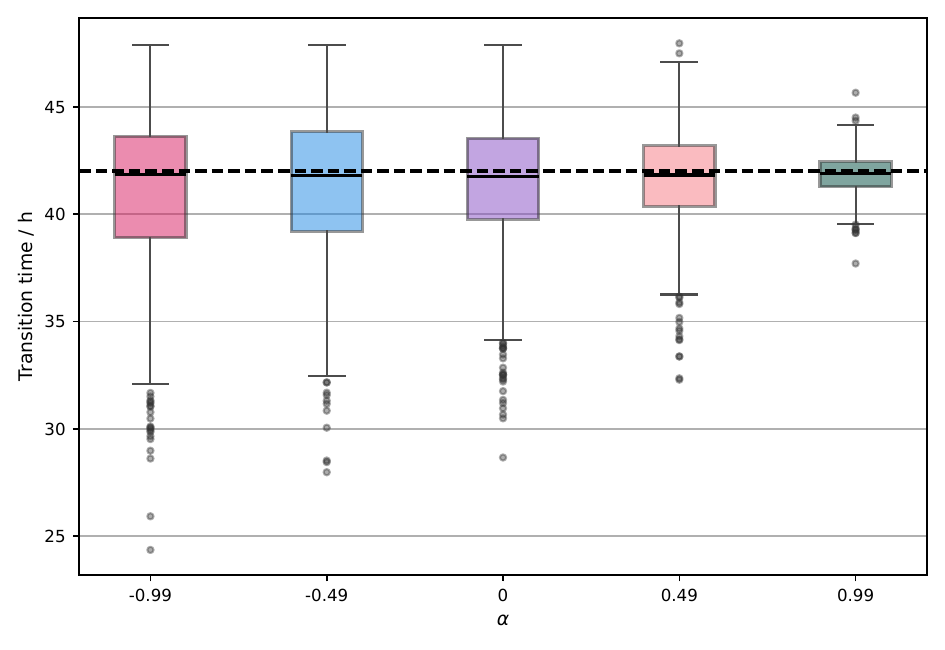}
\caption{As in main text Fig.~\ref{fig:3}A but for noise correlation $\alpha$.}
\end{figure}

\clearpage
\newpage

\bibliographystyle{unsrt}
\bibliography{references}

\end{document}